\documentstyle[aps,prl,multicol,epsf]{revtex}
\def\beginwide{
        \end{multicols} \vspace*{-0.5cm} \noindent
        \rule{3.5in}{.1mm}\rule{.1mm}{5mm} \widetext \medskip }
\def\beginwidetop{
        \end{multicols} \vspace*{-0.5cm} \noindent
        \widetext \medskip }
\def\endwide{
        \hspace*{3.35in}~\rule[-5mm]{.1mm}{5mm}\rule{3.5in}{.1mm}
        \begin{multicols}{2} \vspace*{-1.0cm} \noindent }
\def\endwidebottom{
        \begin{multicols}{2} \vspace*{-1.0cm} \noindent }

\draft

\newcommand{\cT}{\Sigma}
\begin{document}

\title{Exit time of turbulent signals: a way to detect the
intermediate dissipative range.}  

\author{L. Biferale$^{1}$,
M. Cencini$^2$, D.Vergni$^2$ and A. Vulpiani$^{2}$} 

\address{$^1$ Dipartimento di Fisica and INFM, Universit\`a di Roma ``Tor Vergata'',
Via della Ricerca Scientifica 1, I-00133 Roma, Italy  \\
$^2$Dipartimento di Fisica and INFM, Universit\`a di Roma ``La Sapienza'', P.le A. Moro
2, I-00185, Rome, Italy} 

\maketitle
\begin{abstract}
The exit time statistics of experimental turbulent data is analyzed.
By looking at the exit-time moments (Inverse Structure Functions) it is
possible to have a direct measurement of scaling properties of the
laminar statistics.  It turns out that the Inverse Structure Functions
show a much more extended Intermediate Dissipative Range than the 
Structure Functions, leading to the first clear evidence of the
universal properties of such a range of scales.
\end{abstract}

\begin{multicols}{2}

In stationary isotropic turbulent flows,
 a net flux of energy establishes in the 
inertial range, i.e. from forced scales, $L_0$,  down to the
dissipative scale, $r_d$. 
Energy is transferred through a statistically scaling-invariant process,
characterized by a strongly non-gaussian (intermittent) activity.
Understanding the statistical properties of intermittency is
one of the  most  challenging open problems in three dimensional fully
developed turbulence.
In isotropic turbulence, the most studied 
statistical indicators are the longitudinal
structure functions, i.e. moments of the velocity 
increments at distance $R$ in the direction of ${\bf \hat{R}}$: 
\begin{equation}
S_p(R) = <\!\!\left[ ({\bf v}({\bf x+R}) -{\bf v}({\bf x}))\cdot {\bf \hat{R}} \right]^p\!\!>\,.
\label{sf}
\end{equation}
Typically, one is forced to 
analyze one-dimensional string of data: the output
of hot-wire anemometer. In these cases  Taylor
Frozen-Turbulence Hypothesis is used  in order to bridge 
measurements in space with measurements in time. 
Within the Taylor Hypothesis, one 
has the large-scale typical time, $T_0 = L_0/U_0$, and the dissipative
time, $t_d = r_d/U_0$, where $U_0$ is the large scale velocity field,
$L_0$ is the scale of the energy injection and 
$r_d$ is the Kolmogorov dissipative scale. 
As a function of time increment, $\tau$, structure functions  assume the form:
$
S_p(\tau) = <\!\!\left[ (v(t+\tau) -v(t)\right]^p\!\!>$.
It is well know that for time increment corresponding
to  the inertial range, $\tau_d \ll \tau \ll T_0$,
 structure functions develop an anomalous scaling behavior: 
$S_p(\tau) \sim \tau^{\zeta(p)}$, 
where $\zeta(p)$ is a non linear function,
while far inside the dissipative range,
$\tau \ll \tau_d$,  they must show
the laminar  scaling: 
$S_p(\tau) \sim \tau^p$.\\ 
 Beside the huge amount of theoretical,
experimental and numerical studies  devoted to the understanding
of velocity fluctuations  in the inertial range (see \cite{frisch}
for a recent overview), only  few -mainly theoretical- attempts have 
focused on the Intermediate Dissipation Range  (IDR) 
\cite{fv91,jpv,gagne,physicad,procaccia}.
By IDR we  mean the range of scales, $\tau \sim  \tau_d $,
in between the two  
power law ranges:  the inertial and the dissipative range.  \\
The very existence  of the IDR
is relevant for the understanding of many theoretical and
practical issues. Among them we cite: 
the modelizations of small scales for optimizing 
Large Eddy Simulations;  the possible influence of small scales
statistics on macroscopic global quantities, e.g. drag-reduction 
due to the presence of very diluted polymers in the fluid \cite{degennes};
the validity of the Refined Kolmogorov Hypothesis, i.e. 
the bridge between inertial and dissipative statistics.\\
A non-trivial
IDR is connected to the presence 
of intermittent fluctuations in the inertial range. 
Namely, anomalous scaling law characterized by the 
exponents $\zeta(p)$, 
can be explained by assuming  that velocity fluctuations in the inertial
range are  characterized by a spectrum of different local scaling exponents:
$ \delta_{\tau} v = v(t+\tau) -v(t) \sim \tau^{h} $ with the probability to
observe at scale $\tau$ a  value $h$    given by 
$ P_{\tau}(h) \sim \tau^{3-D(h)}$. This is the celebrated multifractal picture
of the energy cascade which has been  confirmed  by many independent
experiments \cite{frisch}. The non trivial 
dissipative statistics  can be explained by defining the dissipative
cut-off as the scale where the local Reynolds number is order of unity:
\begin{equation}
 Re(\tau_d) = \frac{\tau_d\, v_{\tau_d}}{\nu} \sim O(1)\,.
\label{re}
\end{equation}
 By inverting (\ref{re})  
 we immediately obtain a prediction of a fluctuating $\tau_d$:
$$ \tau_d(h)  \sim \nu^{1/(1+h)}\,, $$ 
where for  sake of simplicity
 we have assumed the large scale velocity, $U_0$, and the outer scale,
$L_0$, both fixed to one.  \\
In this letter we propose, and measure in experimental and synthetic data,
a set of new observable which are able to highlight the 
IDR properties. The main idea is to take  a 
one-dimensional string of turbulent data, $v(t)$, 
 and to analyze the statistical
properties of the exit times from a set of defined velocity-thresholds.
 Roughly speaking a kind of {\it Inverse} Structure Functions.\\ 
This analysis allow us to give the first clear evidence of
 non-trivial intermittent fluctuations of the dissipative cut-off
in turbulent signals. 
A similar approach has already been exploited  for  studying the particle
separation statistics \cite{angeloetal}.  Recently, exit-time moments
have also  been  studied in the  time  evolution
of a shell model \cite{mogens}.\\
The letter is organized as follows. First we define the exit-time 
probability density function and we motivate why  this
PDF is dominated by the IDR. Then, we present the 
data analysis performed in high-Reynolds number turbulent flows and in 
synthetic multi-affine signals \cite{bbccv}.  Finally, we  summarize the 
evidences supporting  a non-trivial IDR and  discuss
 possible further investigations. \\
Fluctuations of viscous cut-off are particularly important for all
those regions in the fluid where the velocity field is locally
smooth, i.e. the local fluctuating Reynolds number is small. In this
case, the matching between  non-linear and  viscous terms 
happens at scales  much larger than the Kolmogorov scale,
$\tau_d \sim \nu^{-3/4}$. It is natural, therefore, to look 
for observable which feel  mainly laminar events. 
A possible choice is to measure the {\it exit-time} 
moments through a set of velocity thresholds. 
More precisely, given a reference initial time $t_0$
with velocity $v(t_0)$,  we define $\tau(\delta v)$ as the
first time necessary to have  an absolute  variation equal
 to $\delta v$ in the velocity data, i.e. $|v(t_0)-v(t_0 +\tau(\delta v))| = \delta v$. 
By scanning the whole
time series we  recover  the probability density functions of
$\tau(\delta v)$ at varying $\delta v$ from the typical large scale values down to  the smallest dissipative values.
Positive moments of $\tau(\delta v)$ are   dominated by 
events with a smooth velocity field, i.e. laminar bursts in
the turbulent cascade.  Let us define the 
Inverse Structure Functions (Inverse-SF) as:
\begin{equation}
\cT_{p}(\delta v) \equiv < \!\!\tau^p(\delta v)\!\!> \,.
\label{inv_sf}
\end{equation}
According to the multifractal description
we suppose that, for 
velocity thresholds corresponding to inertial range values of the velocity
differences, $ \delta_{\tau_d}v \equiv v_{m} \ll \delta v \ll v_{M} \equiv \delta_{T_0}v$, 
the following dimensional relation is valid:
\begin{equation}
\delta_{\tau} v \sim \tau^h \;\;\rightarrow \;\; \tau(\delta v) \sim
 \delta v^{1/h} \,,
\end{equation}
with a probability to observe  a value $\tau$  for the exit time
 given by inverting the multifractal probability, i.e. 
\begin{equation}
P(\tau \sim \delta v^{1/h} ) \sim \delta v^{[3-D(h)]/h} 
\end{equation}
Made this ansatz, one can  write down a prediction for the
Inverse-SF, $ \cT_p(\delta v)$ evaluated for  velocity thresholds
within the inertial range:
\begin{equation}
 \cT_p(\delta v) \sim \int_{h_{min}}^{h_{max}} dh \; 
\delta v^{[p + 3-D(h)]/h}  \sim \delta v^{\chi(p)} 
\label{multi}
\end{equation}
where the RHS has been obtained by a saddle point estimate~:  
\begin{equation}
\chi(p) = \min_h\left\{[p + 3-D(h)]/h\right\}\,.
\label{saddle}
\end{equation}
Let us now consider the  IDR properties. \\
For each $p$ the saddle point evaluation (\ref{saddle}) 
selects a particular $h=h_{s}(p)$ where the minimum
is reached. Let us also remark that from (\ref{re}) we have an
estimate for the minimum value assumed by the velocity 
in the inertial range given a certain singularity $h$:
$v_m(h) = \delta_{\tau_d(h)} v \sim \nu^{h/(1+h)}$. 
Therefore, the smallest velocity value at which the scaling 
$\cT_p(\delta v) \sim \delta v^{\chi(p)}$ still holds
depends on both $\nu$ and $h$. Namely,
$\delta v_{m}(p) \sim \nu^{h_{s}(p)/1+h_{s}(p)}$. The most
important  consequence is that for $\delta v < \delta v_{m}(p)$ the integral
(\ref{multi}) is not any more dominated by the saddle point value but
by the maximum $h$ value still dynamically alive at that velocity difference,
$1/h(\delta v) = -1 -\log(\nu)/\log(\delta v)$. 
This leads for $ \delta v < \delta v_{m}(p)$ to  a pseudo-algebraic law:
\begin{equation}
\cT_p(\delta v) \sim
 \delta v^{{\textstyle [p+3-D(h(\delta v))]/h(\delta v)}}\,.
\label{idr}
\end{equation}
The presence of this $p$-dependent velocity range, 
intermediate between the inertial range,
$\cT_p(\delta v) \sim \delta v^{\chi(p)}$, and the far dissipative
scaling, $\cT_p(\delta v) \sim \delta v^{p}$, is the IDR signature. 
Then, it is easy to show that Inverse-SF should display an enlarged IDR.
 Indeed, for the usual {\it direct} 
structure functions (\ref{sf}) the saddle point $h_{s}(p)$ value is reached 
for $h<1/3$. This pushes the IDR to a
range of scales very difficult to observe experimentally \cite{gagne}.
On the other hand, as regards  the Inverse-SF, the saddle point estimate of
positive moments is always reached for $h_{s}(p) > 1/3$.  
This is an indication that we are probing 
the laminar part of the velocity statistics.
Therefore, the presence of the IDR must be felt much earlier
in the range of available velocity fluctuations. 
Indeed, if $h_{s}(p) > 1/3$, the typical velocity field at which the IDR 
shows up is given by $\delta v_{m}(p) \sim \nu^{h_{s}(p)/(1+h_{s}(p))}$, 
that is much larger than the Kolmogorov value $\delta v_{r_d} \sim \nu^{1/4}$. 
\begin{figure}
\narrowtext
\epsfxsize=8truecm
\epsfysize=5truecm
\epsfbox{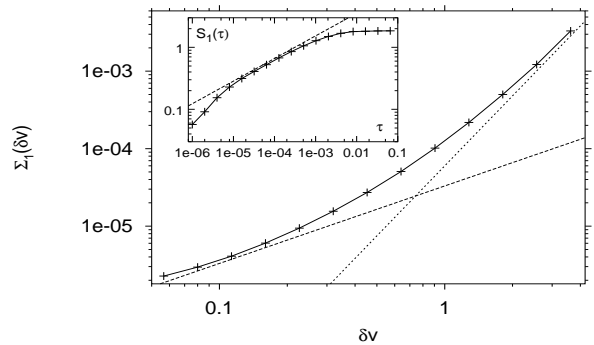}
\caption{Inverse Structure Functions $\cT_1(\delta v)$.
The straight lines shows the dissipative 
range behavior (dashed) $\cT_1(\delta v) \sim  \delta v$, 
and the inertial range non intermittent behavior (dotted)
$\cT_1(\delta v)\sim  (\delta v)^3$. 
The inset shows the direct structure function $S_1(\tau)$ 
with superimposed the  intermittent slope $\zeta(1)=.39$.}
\end{figure}
In Fig. 1 we plot  $\cT_1(\delta v)$ 
evaluated  on a string of high-Reynolds number experimental data as a function
of the available range of velocity thresholds $\delta v$. This data set has been measured
in a wind tunnel  at $Re_{\lambda} \sim 2000$.  \\
Let us first make a technical remark. If one wants to compare the
predictions (\ref{multi}) and (\ref{idr}) with the experimental  data, 
it is necessary to perform the average over
the time-statistics in a weighted way. This is due to the fact that by looking
at the exit-time statistics we are not sampling the time-series
uniformly, i.e. the higher the value of $\tau(\delta v)$ is, 
the longer it is  detectable in the time series.
Let us call $\tau_1(\delta v),\tau_2(\delta v),\dots, \tau_N(\delta v)$ the 
string of exit time values obtained by analyzing the velocity string data
consecutively for a given $\delta v$. $N$ is the number of times for which
$\delta_{\tau}v$ reaches a given threshold. 
It is easy to realize \cite{fsle} that the sequential time average of
any observable based on exit-time statistics, 
$<\!\!\tau^p(\delta v)\!\!>_t \equiv (1/N)  \sum_{i=1}^N \tau_i^p$,  is
connected to the uniformly-in-time multifractal  average, $ < (\cdot) > 
\equiv \int dh (\cdot)$, by the 
relation:
\begin{equation}
<\!\!\tau^p(\delta v)\!\!> =
\sum_{i=1}^{N}  \tau_i^p {\tau_i \over \sum_{j=1}^N \tau_i}
=\frac{<\!\!\tau^{p+1}\!\!>_t}{<\!\!\tau\!\!>_t}\,,
\end{equation} 
where  $\tau_i/\sum_{j=1}^N \tau_i$
 takes into account the non-uniformity in time. 
Let us now go back to Fig. 1. One can see that the scaling is very poor. 
Indeed, it is not possible to extract any quantitative
prediction about the inertial range slope. For this reason, 
we have only drawn the dimensional non-intermittent slope
and the dissipative slope  as a possible qualitative references. 
On the other hand, (inset of Fig, 1),  the 
scaling behavior of the direct structure functions 
$<\!|\delta v (\tau)|\!> \sim \tau^{\zeta(1)}$ 
is quite clear in a wide range of scales. 
This is a clear evidence of  IDR's contamination  into the 
whole range of available velocity values for the Inverse-SF cases.
 Similar results (not shown)
are found for higher orders $\cT_p$ structure functions. 
\begin{figure}
\narrowtext
\epsfxsize=8truecm
\epsfysize=5truecm
\epsfbox{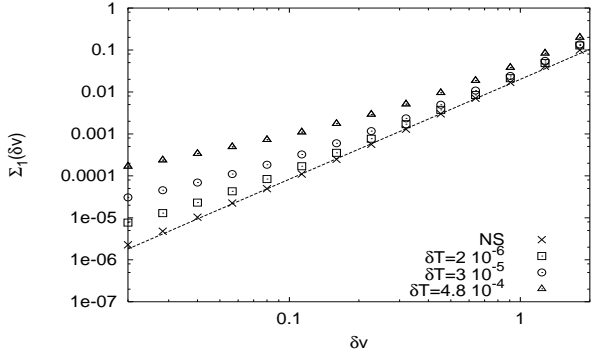}
\caption{
Inverse-Structure-Function $\Sigma_1(\delta v)$ versus $\delta v$ 
for the synthetic signals not
smoothed ($NS$) and smoothed with time windows: 
$\delta T=4.8\cdot 10^{-4},\;3\cdot 10^{-5},\; 2\cdot 10^{-6}$, 
the straight line slope is obtained from the inverse multifractal 
prediction (\ref{saddle}).}
\end{figure}
In order to better understand the scaling properties 
of $\cT_p(\delta v)$ we investigate a synthetic multi-affine 
field  obtained  by combining  successive multiplications
of Langevin dynamics \cite{bbccv}.
The advantage of using  a synthetic  field is that
one can control analytically the scaling properties
of direct structure functions in order to have
the same scaling laws observed in experimental data. 
The signal we used is  sequential in time. Therefore, it  
does not present a superposed hierarchical structure as other  
multi-affine field proposed in the past \cite{physicad93,sreene}.
An IDR can be introduced in the synthetic signals 
by smoothing the original dynamics on a moving time-window
of size $\delta T$.
Imposing a smoothing time-window is  equivalent to fixing 
the  Reynolds numbers,  $Re \sim \delta T^{-4/3}$.
The purpose to introduce this stochastic multi-affine
field is twofold. First we want to reach high Reynolds numbers enough 
to test the inverse-multifractal formula (\ref{saddle}). 
Second, we want  to test that the very extended IDR 
observed in the experimental data,  see Fig. 1,  is also observed
in this stochastic field.  This would  support  the claim that 
the experimental result is the evidence  of an extended IDR. \\
In Fig. 2 we show the Inverse-SF, $\cT_1(\delta v)$, 
measured in the multiaffine synthetic signal at high-Reynolds 
numbers. The observed scaling exponent, 
$\chi(1)$, is in  agreement with the prediction (\ref{saddle}).
The same  agreement also holds  for higher moments.
In Table 1, we compare the best fit to the $\cT_p(\delta v)$ measured
on the synthetic field with the inversion formula (\ref{saddle}). As for
the comparison between the theoretical expectation (\ref{saddle}) 
and the synthetic data let us  note the following points. 
First, in \cite{bbccv} it was
proved that the signal possesses the given direct-structure functions
exponents for positive moments, i.e. the $\zeta(p)$ exponents
are in a one-to-one correspondence with the $D(h)$ curve 
for  $h<1/3$. Nothing was possible to be proved
for observables feeling the $h>1/3$ interval and therefore the agreement
between the inversion formula (\ref{saddle}) and the numerical results
cannot be proved analytically.
Second,  because the synthetic signal is 
defined by using  Langevin processes, the less singular $h$-exponents 
expected to contribute to the saddle-point (\ref{saddle})
is $h=0.5$.  Therefore,  the theoretical prediction, $\chi_{th}(q)$,
in Table 1 has been obtained by imposing $h_{max}=0.5$.  

Let us now go back to the most interesting question about the
statistical properties of the IDR. 
In order to study this question we have smoothed the 
stochastic field, $v(t)$,  by performing a running-time average
over a time-window, $\delta T$. 
Then we compare Inverse-SF scaling properties 
at varying Reynolds numbers, i.e. for different  
dissipative cut-off~: $Re \sim \delta T^{-4/3}$.\\
The expression (\ref{idr}) predicts the possibility
to obtain a data collapse of all curves with 
different Reynolds numbers by rescaling  the Inverse-SF as
follows \cite{fv91,jpv}: 
\begin{equation}
- \frac{\ln (\cT_p(\delta v))}  {\ln (\delta T / \delta T_0)} \;\;vs. \;\; 
-\frac{\ln(\delta v/U)}{\ln (\delta T / \delta T_0)}\,,
\label{rescaling}
\end{equation}
where $U$ and  $\delta T_0$ are adjustable dimensional parameters.
Within the same experimental (or synthetic) 
set up they are Reynolds number independent (i.e. $\delta T$ independent). 
\begin{figure}
\narrowtext
\epsfxsize=8truecm
\epsfysize=5truecm
\epsfbox{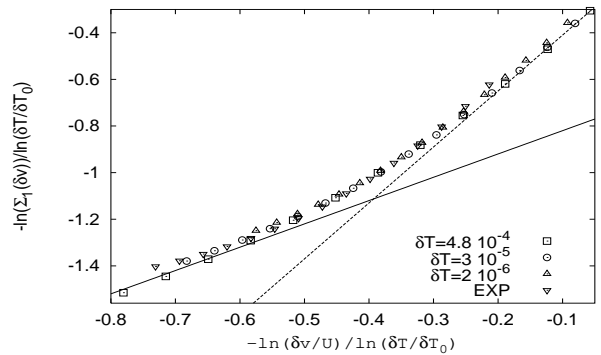}
\caption{Data collapse of the Inverse-SF, $\cT_1(\delta v)$,
 obtained by the rescaling (\ref{rescaling}) for the
smoothed synthetic signals (with time windows: 
$\delta T=4.8\cdot 10^{-4}\,,\;3\cdot 10^{-5}\,,\; 2\cdot 10^{-6}$) 
and the experimental
data ($EXP$). The two straight lines have the dissipative (solid line)
and the  inertial range (dashed) slope.}
\end{figure}
The rationale for the rescale (\ref{rescaling})
stems from the observation that, in the IDR, 
$h_{s}(p)$ is a function of $ \ln(\delta v)/\ln(\nu)$ only.
Therefore, 
identifying $Re \propto \nu^{-1}$,
the relation (\ref{rescaling}) directly follows from
(\ref{idr}). This rescaling was originally proposed as a possible
test of IDR for direct structure functions in \cite{fv91}
but, as already discussed above, for the latter observable
it is very difficult to detect any IDR  due to  the
extremely small  scales involved \cite{gagne}.  \\
Fig. 3  shows the rescaling (\ref{rescaling}) 
of the Inverse-SF, $\cT_1(\delta v)$,  
for the synthetic  field at different Reynolds numbers and 
for the experimental signals.
As it is possible to see, the data-collapse is very good for both the synthetic
and experimental signal. This is a clear evidence that
the poor scaling range observed in Fig. 1
for the experimental signal can be explained as the signature of the IDR. 
The same behavior holds for higher moments (not shown).\\
It is interesting to remark that for a self-affine signal
($D(h)=\delta(h-1/3)$), the IDR is highly reduced and the Inverse-SF, 
scaling trivially as  $ \Sigma_p(\delta v) \sim (\delta v)^{3p} $, 
do not bring any new information. 

Let us summarize the results obtained and the open problems.  First,
by defining the exit-time moments, $\cT_p(\delta v)$,
we argued that they must be dominated by the laminar part of the
energy cascade. This implies that they depend only on the part of
$D(h)$ which falls to the right of its maximum , i.e. $h >
1/3$. These $h$'s values are not testable by the direct structure
functions (\ref{sf}).  Inverse-SF are the natural tool to
test any model concerning velocity fluctuations less singular than the
Kolmogorov value $\delta v \sim \tau^{1/3}$. \\
Second, by
analyzing high-Reynolds data and synthetic fields, we have proved 
 that the extension
of the IDR for $\cT_p(\delta v)$ is magnified. Moreover, the
rescaling (\ref{rescaling}) based on the assumption (\ref{re}) gives a
good data collapse of all curves for different Reynolds numbers. This
is a clear evidence of the IDR.\\
Many questions are still open.
First, the analysis of a wider set of experimental data could make it possible
 to quantify the agreement of the data-collapse with the prediction based
on (\ref{re}) and (\ref{idr}).  Indeed, it is easy to realize that, by
using different parameterization for the onset of the viscous range, one
would have predicted the existence of an extended IDR for
$\cT_p(\delta v)$ but with a slightly different rescaling
procedure \cite{physicad}.  The quality of experimental data available
to us is not high enough to distinguish between the two different
predictions.  Analyzing different experimental data-sets, at different
Reynolds numbers, could also make it possible 
to better explore $D(h)$ for $h >
1/3$.  This is an important question because doubts about 
the universality of these $D(h)$  values may  be raised on
the basis  of the usual energy cascade picture.  For example, 
as discussed above, in the Langevin synthetic-data a good agreement between
the multifractal prediction and the numerical data 
is obtained  by imposing $h_{max}=0.5$, similarly in true turbulent data
other $h_{max}$ values could appear depending on the physical mechanism
driving the energy transfer at large scales. \\
Once the attention is focused on the exit-time
statistics, different questions connected to the entropic
properties of the exit-times can also be asked.  For these  kind of
questions there are no {\it a priori} reasons to believe that the
information coded in the direct-statistics is similar to the
information coded in the inverse-statistics.  Work is in progress in
this direction.

We acknowledge useful discussions with R. Benzi, G. Boffetta, A. Celani, 
M.H. Jensen, P. Muratore Ginanneschi,  M. Vergassola. We also thank Y. Gagne for the access 
to the  experimental data. 
This work has been partially supported by INFM (PRA-TURBO) and by 
the European Network {\it Intermittency in Turbulent Systems} 
(contract number FMRX-CT98-0175).

\begin{table}[htb]
\caption{Comparison between the Inverse-SF scaling exponents $\chi_{syn}(p)$
measured in the synthetic signal and the inversion of the theoretical 
multifractal prediction (\ref{saddle}), $\chi_{th}(p)$. 
The synthetic signal has been defined such has the $D(h)$ function 
leads to the same set of experimental $\zeta(p)$ exponents for the 
direct structure functions.}
\vspace*{2mm}
\begin{tabular}{||c|c|c|c|c|c||} 
p  & 1 & 2 & 3 & 4 & 5 \\
\hline
$~\chi_{syn}(p)~$ & 2.32(4)& 4.40(8) & 6.38(8)& 8.3(1) & 10.1(2) \\
$~\chi_{th}(p)~$ & 2.32 & 4.34 & 6.34& 8.35 & 10.35\\
\end{tabular}
\end{table}
\end{multicols}

\begin{thebibliography}{99}
\bibitem{frisch}U. Frisch, {\it Turbulence. The legacy of A.N. Kolmogorov}, 
Cambridge University Press, Cambridge (1995). 
\bibitem{fv91} U. Frisch and M. Vergassola, Europhys. Lett {\bf 14}
439 (1991).
\bibitem{jpv}M.H. Jensen, G. Paladin and A. Vulpiani, Phys. Rev.
Lett. {\bf 67} 208 (1991).
\bibitem{gagne} Y. Gagne and B. Castaing, C. R. Acad. Sci. Paris
{\bf 312} 441 (1991).
\bibitem{physicad}R. Benzi, L. Biferale, S.Ciliberto, M.V. Struglia 
and R.Tripiccione, Physica D, {\bf 96}, 162 (1996). 
\bibitem{procaccia} V.S. L'vov and I. Procaccia,
Phys. Rev. E, {\bf 54}, 6268 (1996). 
\bibitem{degennes} M. Tabor and P.G. de Gennes, Europhys. Lett
{\bf 2} 519 (1986).
\bibitem{angeloetal} G. Boffetta, A. Celani, A. Crisanti and
 A. Vulpiani, ``Relative dispersion in fully developed turbulence:
 Lagrangian statistics in synthetic flows'',
 Europhys. Lett. in press (1999) chao-dyn/9803030.
\bibitem{mogens} M.H. Jensen,  ``Multiscaling and 
Structure Functions in Turbulence: An Alternative Approach'' chao-dyn/9901021.
\bibitem{bbccv} L. Biferale, G. Boffetta, A. Celani, A. Crisanti and 
A. Vulpiani, Phys. Rev. E {\bf  57}  R6261 (1998). 
\bibitem{fsle} E. Aurell, G. Boffetta, A. Crisanti, G. Paladin and A. Vulpiani,
Phys. Rev. Lett. {\bf 77} 1262 (1996); J. Phys. A {\bf 30} 1 (1997) 
\bibitem{physicad93} R. Benzi, L. Biferale, A. Crisanti, 
G. Paladin, M. Vergassola, A. Vulpiani, Physica D {\bf 65}, 352 (1993). 
\bibitem{sreene}
  A. Juneja, D. P. Lathrop, K. R. Sreenivasan, and G. Stolovitzky, 
  Phys. Rev. E {\bf  49}, 5179 (1994).
\end{thebibliography}
\end{document}